%% file: ms.tex
\pdfoutput=1
\documentclass[aip,jcp,reprint,floatfix,superscriptaddress]{revtex4-1}

\usepackage{amsmath,amsfonts,amssymb} %
\usepackage{mathtools} %
\ifx\phone\undefined
\usepackage{wasysym} 
\else
\let\Phone\phone
\let\phone\relax
\usepackage{wasysym}

\let\phone\Phone
\let\Phone\relax
\fi
\usepackage{bm} %
\usepackage[bbgreekl]{mathbbol} %
\usepackage{graphicx} %
\usepackage[dvipsnames,table]{xcolor} %
\usepackage{multirow,tabularx} %
\usepackage[acronym]{glossaries} %
\usepackage{braket} %
\usepackage[binary-units=true,detect-weight=true]{siunitx} %
\usepackage{fancyhdr} %
\usepackage{microtype} %
\usepackage{natmove} 
\usepackage[caption=false]{subfig} %
\usepackage[breaklinks, %
colorlinks, linkcolor=Blue, citecolor=Blue, urlcolor=Blue]{hyperref} %
\usepackage[version=3]{mhchem} %
\usepackage{wrapfig} %
\AtBeginDocument{\usepackage{booktabs}}
\usepackage[linesnumbered,lined,boxed,ruled]{algorithm2e}

\DeclareSIUnit[number-unit-product = {\,}]\cal{cal}
\DeclareSIUnit\erg{erg}
\DeclareSIUnit\au{au}
\def\MyTitle{Thawed Gaussian wave packet dynamics: a critical
  assessment of three propagation schemes} %
\def\MyAuthora{Ilya G. Ryabinkin} %
\def\MyAuthorb{Rami Gherib} %
\def\MyAuthorc{Scott N. Genin} %

\def\MySubject{} %

\hypersetup{%
  pdftitle={\MyTitle{}}, %
  pdfauthor={\MyAuthora{}, \MyAuthorb{}, and \MyAuthorc{}}, %
  pdfsubject={\MySubject{}}, %
  pdfkeywords={} %
}

\newcolumntype{Y}{>{\centering\arraybackslash}X}

\input{acronyms}

\input{operators}

\begin{document}
\ifx\latin\undefined %
  \newcommand{\latin}{\textit} %
\fi%

\title{\MyTitle}

\author{\MyAuthora{}} %
\email{ilya.ryabinkin@otilumionics.com} %

\author{\MyAuthorb{}} %

\author{\MyAuthorc{}} %
\affiliation{OTI Lumionics Inc., 3415 American Drive Unit 1, \\
  Mississauga, Ontario L4V\,1T4, Canada} %


\date{\today}

\begin{abstract}
  We assessed three schemes for propagating a variable-width (thawed)
  Gaussian wave packet moving under the influence of Morse or
  double-well potentials with parameters that are chemically
  representative. The most rigorous scheme is based on the \gls{TDVP};
  it leads to realistic behaviour of the center and width of a wave
  packet in all investigated regimes. Two other approximate schemes,
  Heller's and the extended semiclassical ones, demonstrate various
  aberrations. Heller's scheme does not properly account for various
  \acrlong{ZPE}-related effects, is unable to predict tunneling, and
  more importantly, exhibits completely nonphysical unbound width
  oscillations. The extended semiclassical scheme, which was developed
  to address some of the shortcomings of the Heller counterpart,
  demonstrates another unphysical behaviour: self-trapping of a
  trajectory in both Morse and double-well potentials. We conclude
  that only the \gls{TDVP}-based scheme is suitable for problem-free
  dynamical simulations. This, however, raises the question of how to
  utilize it efficiently in high-dimensional systems.
\end{abstract}

\glsresetall

\maketitle

\section{Introduction}
\label{sec:introduction}

An essential part of theoretical modelling of novel fluorescent or
phosphorescent materials for the display industry is simulation of
their vibrationally resolved (vibronic) optical spectra. They are
naturally accessible \latin{via} the Fourier transform of a
dipole-dipole autocorrelation function
$C(t)$~\cite{Heller:1978/jcp/2066, Heller:1981/acr/368,
  Mukamel:1982/jcp/173, Baiardi:2013/jctc/4097},
\begin{equation}
  \label{eq:autocorrelation_function}
  C(t) = \Braket{\Psi(0)|\Psi(t)}
\end{equation}
where $\Psi(0) = \mu_{fi}\chi_i$, is a product of nuclear wave
function of the initial state $\chi_i$ and a transition dipole moment
function $\mu_{fi}$ associated with an $i \to f$ electronic
transition,
\begin{equation}
  \label{eq:final_state}
  \Psi(t) = \exp(-\I {\hat H}_f t) \Psi(0)
\end{equation}
is a time-dependent wave function that evolves under the action of a
nuclear Hamiltonian ${\hat H}_f$. Thus, the problem is reduced to
exact or approximate quantum evolution of some initial wave function.

In the past decades, several exact methods for solving the \gls{TDSE}
have been developed: the split-operator
method~\cite{Fleck:1976/ap/129, Feit:1982/jcp/412, Feit:1983/jcp/301,
  Feit:1984/jcp/2578, Bandrauk:1992/cjc/555}, the Chebyshev polynomial
expansion method~\cite{TalEzer:1984/jcp/3967,
  Balakrishnan:1997/pr/79}, \gls{MCTDH}~\cite{Meyer:1990/cpl/73,
  Wang:2003/jcp/1289}, and \gls{vMCG}~\cite{Richings:2015/irpc/269}
methods, to name a few. However, exact methods in general scale
exponentially with the number of quantum \glspl{DOF} and, are thus
unsuitable for molecules of the size of typical light emitters.

Vibronic spectra simulations, on the other hand, rarely require
accurate long-time dynamics. As a result, \emph{approximate} quantum
dynamics methods with better computational scaling and hence
applicable to larger systems received a lot of attention. One of the
oldest yet still very popular approach uses a single Gaussian wave
packet to represent a nuclear wave
function~\cite{Heller:1975/jcp/1544, Lee:1982/jcp/3035,
  Herman:1984/cp/27, Arickx:1986/cpl/310, Singer:1986/mp/761,
  Coalson:1990/jcp/3919, Pattanayak:1994/pre/3601, Ando:2003/cpl/532,
  Ando:2004/jcp/7136, Worth:2004/fd/307, Faou:2006/cvs/45,
  Burghardt:2008/jcp/174104,HyeonDeuk:2009/jcp/064501,
  Ohsawa:2013/jpamt/405201, Skouteris:2014/jcp/244104,
  Coughtrie:2015/jcp/044102, Gu:2016/jpca/3023, Vacher:2016/tca/1,
  JoubertDoriol:2017/jpcl/452, Patoz:2018/jpcl/2367,
  Han:2018/jctc/5527, Polyak:2019/jcp/041101}. The exact time
evolution of a wave function is replaced by time evolution of Gaussian
parameters; to derive the corresponding \glspl{EOM} different guiding
rules may be applied. In this paper we consider the three most popular
schemes for the Gaussian propagation: Heller's scheme, also known as
the \gls{LHA}~\cite{Heller:1975/jcp/1544}, its correction---the
extended semiclassical dynamics~\cite{Pattanayak:1994/pre/3601,
  Ohsawa:2013/jpamt/405201}, and the full, most rigorous scheme based
on the \gls{TDVP}\cite{Book/Kramer:1981, Kan:1981/pra/2831}, which was
independently derived by many authors~\cite{Rajagopal:1982/pra/2977,
  Corbin:1982/mp/671, Heather:1985/cpl/558, Sawada:1985/jcp/3009,
  Singer:1986/mp/761, Arickx:1986/cpl/310, Grad:1987/jcp/3441,
  Coalson:1990/jcp/3919, Pattanayak:1994/pre/3601,
  Burghardt:1999/jcp/2927, Worth:2004/fd/307, Ando:2004/jcp/7136,
  HyeonDeuk:2009/jcp/064501, Skouteris:2014/jcp/244104,
  JoubertDoriol:2017/jpcl/452} though not all of them recognized that
they used one of the equivalent formulations of
\gls{TDVP}~\cite{Broeckhove:1988/cpl/547}.

An early assessment of both approximate schemes---Heller's and the
extended semiclassical ones---was made
by~\citet{Pattanayak:1994/pre/3601} who considered the dynamics of a
Gaussian wave packet placed atop of a barrier at $x=0$ of a quartic
potential, $V(x) = ax^4 - bx^2 + c$, with $a, b >0$. They observed
that the wave packet spread without bounds, which was clearly
unphysical. Qualitative differences between Heller's and the extended
semiclassical schemes were reported in a study of a tunneling escape
from a cubic potential $V(x) = x^2/2 - cx^3$,
$c =1/10$~\cite{Ohsawa:2013/jpamt/405201}. The Heller trajectory
remained trapped, while the semiclassical one escaped through a
barrier. \Citet{Heather:1985/cpl/558} compared the Heller scheme with
their ``minimum error method'', which was a \gls{TDVP}-based scheme in
disguise, and noticed that the former showed substantially larger
deviations from the exact quantum result than the latter, even in the
low-energy regime when oscillations took place near the minimum of the
potential. They speculated that the observed errors could be
attributed to disregarding the Heisenberg's principle and poor
description of the \gls{ZPE} in the Heller approximation. Several
additional issues were documented by the same authors in
Ref.~\citenum{Heather:1986/jcp/3250}. A failure of Heller's scheme in
simulations of liquid argon was noted in
Ref.~\citenum{Corbin:1982/mp/671}. However, in a study of collinear
\ce{ICN} photodissociation~\cite{Coalson:1990/jcp/3919}, Heller's
scheme demonstrated reasonable agreement with more accurate
\gls{TDVP}-based Gaussian dynamics. Heller's scheme coupled with
on-the-fly calculations of \latin{ab initio} molecular energies,
gradients, and Hessians was recently employed in simulations of
optical and photolectron spectra of polyatomic molecules and the
reported outcomes were encouraging~\cite{Wehrle:2014/jcp/244114,
  Wehrle:2015/jpca/5685, Patoz:2018/jpcl/2367}. The full
\gls{TDVP}-based scheme---to the best of our knowledge---was used only
twice in simulations of realistic systems: solid and liquid argon with
the Lennard-Jones potential fitted by a linear combination of
Gaussians~\cite{Corbin:1982/mp/671}, and liquid water employing
analytically parameterised intra- and intermolecular
potentials~\cite{HyeonDeuk:2009/jcp/064501}. In the latter work, the
observed differences in simulated IR and Raman spectra between the
\gls{TDVP}-based scheme and purely classical simulations as well as
better agreement with experimental spectra for the former were
attributed to proper description of essential quantum effects.
Surprisingly, although multiple successful applications and a few
points of failures were reported, a systematic comparative assessment
of all three Gaussian propagation schemes is still missing in the
literature.

In this study, our primary goal is to create a consistent picture of
the successes and failures of the Gaussian propagation schemes. To
this end, we study their performance for one-dimensional (1D) model
anharmonic potentials, ``quantum potentials'' for short, which allows
us to decouple a question of the quality of employed potentials from
consequences of dynamical approximation. Our quantum potentials are
also amenable for extensive analytical treatment, and we complement
our numerical results with analytical considerations. Although it was
entirely possible, we refrained from the comparison to the exact
quantum results as in realistic molecular applications the exact
quantum solution will not be available. Instead, we focus on the
analysis of internal consistency of the resulting dynamics.

Our analysis relies on a unique mapping between 1D \textit{quantum}
dynamics of a thawed Gaussian wave packet and 2D \textit{classical}
dynamics of a fictitious particle of the same mass moving under the
influence of an \emph{extended potential} $U(q,w)$, where $q$ and $w$
are the position of the center and the width of the corresponding
Gaussian~\cite{Singer:1986/mp/761, Arickx:1986/cpl/310,
  Pattanayak:1994/pre/3601}. As we show, $U$ compactly encodes some
features that are commonly referred to as ``quantum'', for example,
\gls{ZPE}~\cite{Miller:1989/jcp/2863, Alimi:1992/jcp/2034,
  Peslherbe:1994/jcp/1179, Ben-Nun:1994/jcp/8768, Sewell:1992/cpl/512,
  Lim:1995/jcp/1705, Guo:1996/jcp/576, Habershon:2009/jcp/244518,
  Gabor:2010/jcp/164103}, so that our analysis extends beyond purely
dynamical perspective.

The paper is organized as follows. First, we formulate the \glspl{EOM}
for all three schemes in a common form using the aforementioned
mapping. Next, we introduce our model potentials---the Morse and the
double-well ones---that are parametrized to represent realistic
physical systems. We analyse a static picture and consider several
dynamical regimes, such as a small-amplitude motion near a minimum,
barrier tunneling/penetration, and large amplitude motion. We assess
not only spectroscopic, but also mechanical properties and we attempt
to identify all cases when the approximate schemes fail and
rationalize why. What is beyond the scope of our study is the
vibrational mode coupling due to anharmonicity as well as
Dushinsky~\cite{Baiardi:2013/jctc/4097} rotations --- both require
genuine multi-dimensional potentials. However, many shortcomings of
approximate schemes are evident even in 1D systems.

\section{Theory}
\label{sec:theory}

\subsection{Thawed Gaussian dynamics as classical dynamics in extended
  phase space }
\label{sec:thaw-gauss-dynam}

An $N$-dimensional Gaussian wave packet ($N$ being the number of
quantum \glspl{DOF} in a problem) is customarily parameterized
as~\cite{Heller:1975/jcp/1544,Corbin:1982/mp/671}
\begin{align}
  \label{eq:gwp_nd}
  \psi(\mathbf{x}, t) & = \nonumber\\
                      & \hspace{-2em} \exp\left\{\frac{\I}{\hbar}\left[(\mathbf{x} -
                        \mathbf{q}_t)^TA_t(\mathbf{x} - \mathbf{q}_t) +
                        \mathbf{p}_t\cdot(\mathbf{x} - \mathbf{q}_t) +
                        \gamma_t\right]\right\},
\end{align}
where $\mathbf{q}_t$ and $\mathbf{p}_t$ are time-dependent coordinates
and momenta of a center of the Gaussian in coordinate and momentum
spaces, respectively, $A_t$ is an $N\times N$ complex symmetric
time-dependent matrix, and $\gamma_t$ is a complex scalar. In what
follows, the case of $N=1$, in which vectors and matrices in
Eq.~\eqref{eq:gwp_nd} become scalars, is exclusively considered. For
brevity, of all parameters are assumed to be time dependent, and the
subscript $t$ is suppressed.

Parametrization of a Gaussian wave packet as in Eq.~\eqref{eq:gwp_nd}
is not physically transparent. While $p$ and $q$ are
quantum-mechanical averages of the position and momentum operators,
respectively,
\begin{align}
  \label{eq:mean_x}
  \braket{\hat x}  & = q, \\ 
  \braket{\hat p}  & = p, 
\end{align}
the physical meaning of $A$ is obscured by its complex-valued nature.
One can separate real and imaginary parts of $A$ and parametrize them
as
\begin{equation}
  \label{eq:A_wu}
  A = \frac{u}{2w} + \I\frac{\hbar}{4w^2},
\end{equation}
so that the one-dimensional Gaussian assumes the
form~\cite{Arickx:1986/cpl/310}
\begin{align}
  \label{eq:1dgwp_uw_unnormalized}
  \psi(x, t) & = \nonumber\\
             & \hspace{-2em} \exp{\left\{-\dfrac{1}{4w^2}\left(1-2\I
               w\frac{u}{\hbar}\right)\left(x-q\right)^2 +
               \frac{\I}{\hbar}p\left(x-q\right) + \frac{\I}{\hbar}\lambda\right\}},
\end{align}
and
\begin{equation}
  \label{eq:width_squared}
  \frac{\int{|\psi(x, t)|^2(x - q)^2\,\mathrm{d}x}}{\int{|\psi(x,
      t)|^2\,\mathrm{d}x}} = w^2,
\end{equation}
giving $w$ the meaning of the width of a wave packet.
$\lambda = \Re{\{\gamma\}}$ is a real parameter, whereas the imaginary
part of $\gamma$ can be absorbed into the norm $\mathcal{N}$,
\begin{equation}
  \label{eq:GWP_norm}
  \mathcal{N}^2 = \braket{\psi|\psi} = \sqrt{2\pi}w.
\end{equation}

Time evolution of the parameters $p$, $q$, $w$, $u$, and $\lambda$ can
be derived in the most general way using the
\gls{TDVP}\cite{Book/Kramer:1981, Kan:1981/pra/2831}. The resulting
\glspl{EOM} are~\cite{Arickx:1986/cpl/310}
\begin{subequations}
  \label{eq:pq_dot}
  \begin{align}
    \label{eq:q_dot}
    {\dot q} = \{q, \mathcal{H}\} & = \frac{p}{m}, \\
    \label{eq:p_dot}
    {\dot p} = \{p, \mathcal{H}\} & = -\frac{\partial U}{\partial q},
  \end{align}
\end{subequations}
\begin{subequations}
  \label{eq:uw_dot}
  \begin{align}
    \label{eq:w_dot}
    {\dot w} = \{w, \mathcal{H}\} & = \frac{u}{m}, \\
    \label{eq:u_dot}
    {\dot u} = \{u, \mathcal{H}\} & = -\frac{\partial U}{\partial w},
  \end{align}
\end{subequations}
where dots symbolize the full time derivative, $\{\cdot, \cdot\}$ are
Poisson brackets and $\mathcal{H}$ is a classical Hamilton function in
a \emph{four}-dimensional phase space of canonically conjugated
variables $(p, q)$ and $(u, w)$ defined by the quantum Hamiltonian
\begin{equation}
  \label{eq:Ham}
  \hat H = -\frac{\hbar^2}{2m}\frac{\mathrm{d}^2}{\mathrm{d} x^2} + V(x)
\end{equation}
as
\begin{align}
  \label{eq:H_func}
  \mathcal{H}(q,p,w,u) & = \mathcal{N}^{-2}\Braket{\psi|\hat H|\psi} = \frac{p^2}{2m} + \frac{u^2}{2m} + U(q,w), \\
  \label{eq:U_func}
  U(q,w) = & \frac{\hbar^2}{8mw^2} +
             \frac{1}{\sqrt{2\pi}}\int\limits_{-\infty}^\infty V(wx + q)
             \E^{-x^2/2}\,\mathrm{d}x.
\end{align}
Thus, quantum dynamics of one-dimensional Gaussian is exactly
represented by classical dynamics of a particle in a phase space whose
dimensionality is twice as large as that of the parent quantum model.
As was mentioned above, $w$ is a width of the packet and $u$ is a
momentum canonically conjugated to it.

The last equation that is relevant to computations of optical spectra
is the \gls{EOM} for the position-independent phase of a wave packet,
$\lambda$. It reads~\cite{Pattanayak:1994/pre/3601}
\begin{equation}
  \label{eq:lambda_dot}
  \dot\lambda = \frac{\dot w u - \dot u w}{2} + p \dot q - E,
\end{equation}
with
\begin{align}
  \label{eq:Edef}
  E = \mathcal{H}|_\text{at the trajectory}.
\end{align}

\subsection{Representations of $U(q,w)$ and approximate dynamical
  schemes}
\label{sec:representations-uq-w}

The downside of a classical view on the thawed Gaussian dynamics is
that $U$ requires evaluation of a parametric integral in
Eq.~\eqref{eq:U_func}. $U(q, w)$ is an even function of $w$;
additionally, if $V$ is analytic, one can
write~\cite{Pattanayak:1994/pre/3601}
\begin{equation}
  \label{eq:Vint_exp}
  \frac{1}{\sqrt{2\pi}}\int\limits_{-\infty}^\infty V(wx + q) \E^{-x^2/2}\,\mathrm{d}x = V(q) +
  \sum_{n=1}^\infty V^{(2n)}(q)\frac{w^{2n}}{2^nn!}. 
\end{equation}
Numerical evaluation of $U(q,w)$ \latin{via} integration or
series~\eqref{eq:Vint_exp} summation is computationally intensive
unless $U$ can be found in a closed form~\cite{Ando:2003/cpl/532,
  Ando:2004/jcp/7136, HyeonDeuk:2009/jcp/064501}.

In order to employ realistic, possibly \latin{ab initio} potentials
$V$, simplified means to compute $U$ are thus highly desirable. The
lowest-order approximations of $U$ that follow from
Eq.~\eqref{eq:Vint_exp} are:
\begin{align}
  \label{eq:U_0}
  U^0(q, w) & = \frac{\hbar^2}{8mw^2} + V(q), \\
  \label{eq:U_1}
  U^1(q, w) & = \frac{\hbar^2}{8mw^2} + V(q) + \frac{w^2}{2} V''(q).
\end{align}
$U^0(q)$ is separable, and dynamics of $(p, q)$ and $(u, w)$ guided by
$U = U^0$ is fully decoupled: a particle of a mass $m$ with a momentum
$p$ and a position $q$ feels a classical force $-V'(q)$, while the
width $w$ grows infinitely under a purely repulsive force
$\hbar^2/4mw^3$. The latter behaviour is consistent with a Gaussian
wave packet moving and spreading on a potential with vanishing second-
and higher-order derivatives (flat potentials) but is less relevant to
molecules. At the next order, the potential $U^1(q, w)$ can be
evaluated in a closed form for potentials $V$ that are at most
quadratic in $x$,
\begin{align}
  \label{eq:U_LHA}
  U^\text{LHA}(q,w) & = \frac{\hbar^2}{8mw^2} + V(q) + \frac{kw^2}{2},
\end{align}
where $k = V''(q) = const$ is a (force) constant. For such potentials
the \glspl{EOM} for the pair $(p, q)$ are still purely classical as
$-\partial U^\text{LHA}/\partial q = -V'(q)$ because the derivative of
a constant term $kw^2/2$ vanishes. The \glspl{EOM} for the $(u, w)$
pair can be converted into a second-order equation,
\begin{align}
  \label{eq:w_LHA}
  m\ddot w = \frac{\hbar^2}{4mw^3} - kw,
\end{align}
and integrated analytically (see, for example,
Ref.~\citenum{Arickx:1986/cpl/310}) to give harmonic oscillations of
$w$ around the initial value $w_0$ with an amplitude that vanishes
when
\begin{equation}
  \label{eq:coherent_state_width}
  w_0 = (\hbar^2/4mk)^{1/4},
\end{equation}
which is a width of a Gaussian \emph{coherent
  state}~\cite{Book/Perelomov:1972, Rajagopal:1982/pra/2977}.

The assumption that the quantum potential $V$ can be replaced by its
quadratic approximation centered at $q$---the \acrfull{LHA}---was
originally considered by~\citet{Heller:1975/jcp/1544}. The resulting
\glspl{EOM}, purely classical for the $(p, q)$ pair and given by
Eqs.~(\ref{eq:w_dot}--\ref{eq:u_dot}) with $U = U^1(q, w)$ for the
pair $(u, w)$, are known as Heller's
dynamics~\cite{Heller:1975/jcp/1544, Pattanayak:1994/pre/3601}. It is
exact for harmonic potentials, but Heller himself suggested to apply
it to \emph{anharmonic} potentials, for which $V''(q)$ is not
constant. One inconsistency of this approach, which was pointed out
by~\citet{Pattanayak:1994/pre/3601} (see also
Ref.~\citenum{Ohsawa:2013/jpamt/405201}), is immediately clear:
\glspl{EOM} for pairs $(p,q)$ and $(u, w)$ employ potentials that
differ by more than a constant if $V(q)$ is anharmonic. To restore
consistency and use, for example, a common $U^1(q,w)$, one must write
\begin{align}
  \label{eq:force_extended_semicl}
  \dot p = -\frac{\partial U^1}{\partial q} & = -V'(q) - \frac{w^2}{2}
                                              V^{(3)}(q).
\end{align}
Thus-modified \glspl{EOM} constitute the ``extended semiclassical
dynamics'' in the terminology of~\citet{Pattanayak:1994/pre/3601}. The
presence of the third derivative of $V$ as well as coupling between
$(p,q)$ and $(u,w)$ pairs make the extended semiclassical dynamics
computationally intensive.

The \gls{EOM} for the phase $\lambda$ is also modified. For both
approximate schemes one must use
\begin{align}
  \label{eq:Edef_heller+ext_semicl}
  E = \frac{p^2}{2m} + \frac{u^2}{2m} + U^1(q,w)
\end{align}
in Eq.~\eqref{eq:Edef}~\cite{Pattanayak:1994/pre/3601}. Note that the
semiclassical dynamics conserves $E$ as defined by
Eq.~\eqref{eq:Edef_heller+ext_semicl} while Heller's dynamics
conserves only classical energy $E_\text{cl} = \frac{p^2}{2m} + V(q)$.
This is another inconsistency in the Heller approach.

\section{Simulations setup}
\label{sec:simulations-setup}

\subsection{Model potentials}
\label{sec:model-potentials}

We employed two one-dimensional anharmonic potentials. The first is
the Morse potential~\cite{Morse:1929/pr/57},
\begin{equation}
  \label{eq:morse_def}
  V_\text{M}(x) = D_e \left[1- \E^{-b(x-a)}\right ]^2 - D_e,
\end{equation}
which describes stretching of an \ce{O-H} bond in certain types of
clays~\cite{Greathouse:2009/jcp/134713}. The reduced mass of a
particle associated with this type of motion is chosen to be a proton
mass, $m \approx \SI{1836}{\electronmass}$. Numerical values of the
parameters $D_e$, $a$, and $b$ are listed in
Table~\ref{tab:quantum_potentials_params}.

The second model potential is a double-well potential for the
inversion motion in ammonia, \ce{NH3}~\cite{Swalen:1962/jcp/1914}. It
is a combination of a harmonic well and a Gaussian barrier at its
center,
\begin{equation}
  \label{eq:ammonia_def}
  V_\text{A}(x) = \frac{1}{2}kx^2 + b \E^{-cx^2},
\end{equation}
Numerical values of the parameters and the reduced mass $m$ for the
inversion motion are given in
Table~\ref{tab:quantum_potentials_params}.
\begin{table}[!b]
  \centering
  \caption{Parameters of the model potentials.}
  \begin{minipage}{1.0\linewidth}
    \begin{tabularx}{1.0\linewidth}{@{}Xrr@{}} 
      \toprule %
      Parameter     & \multicolumn{2}{c}{Value} \\
      \cmidrule{2-3}
                    & {atomic units}  & {common units} \\
      \midrule
                    & \multicolumn{2}{l}{\slshape Morse potential, Eq.~\eqref{eq:morse_def}} \\[0.5ex]

      $D_e$         & \num[round-mode=figures,round-precision=6]{0.2107523558866643} & \SI[]{132.2491}{\kilo\cal\per\mol} \\
      $a$           & \num[round-mode=figures,round-precision=6]{1.7857911877713528} & \SI{0.9450}{\angstrom}           \\
      $b$           & \num[round-mode=figures,round-precision=6]{1.1544000855848946} & \SI{2.1815}{\per\angstrom}       \\
      $m$           & \num[round-mode=figures,round-precision=6]{1836.15267344     } & \SI[round-mode=figures,round-precision=6]{1.0072764666250018}{\dalton}  \\[0.5ex]
                    & \multicolumn{2}{l}{\slshape Ammonia double-well potential, Eq.~\eqref{eq:ammonia_def}} \\[0.5ex]
      $k$           & \num{0.07598}            & \SI{11.83e4}{dyns\per\cm} \\
      $b$           & \num{0.05684}            & \SI{2.478e-12}{\erg} \\
      $c$           & \num{1.3696}             & \SI{4.891e16}{\per\square\cm} \\
      $m$           & \num{4668}               & \SI{2.561}{\dalton} \\
      \bottomrule
    \end{tabularx}
  \end{minipage}
  \label{tab:quantum_potentials_params}
\end{table}

\subsection{Evaluation of the extended potential}
\label{sec:eval-extend-potent}

The extended potential $U(q, w)$ can be found in a closed form for
both model potentials, Eqs.~\eqref{eq:morse_def} and
\eqref{eq:ammonia_def}, respectively:
\begin{align}
  \label{eq:Umorse_an}
  U_\text{M}(u, w) = & \frac{\hbar^2}{8mw^2} \nonumber \\
                     & + D_e\left[\E^{-2b(q-a-bw^2)} - 2\E^{-b(q-a-bw^2/2)}\right], \\
  \label{eq:Uammonia_an}
  U_\text{A}(u, w) = & \frac{\hbar^2}{8mw^2} +
                       \frac{1}{2}k(q^2+w^2) +
                       \frac{b}{\sqrt{z}}\E^{-cq^2/z},
\end{align}
where $z = 1+2cw^2$.

Application of the thawed Gaussian dynamics together with \latin{ab
  initio} potentials requires numerical schemes for computing $U$. We
implemented two such schemes. The first one is applying an $n$-point
Gauss-Hermite quadrature~\cite{Townsend:2015/ijna/337} to the Gaussian
integral in Eq.~\eqref{eq:U_func},
\begin{equation}
  \label{eq:GH_scheme}
  \int\limits_{-\infty}^\infty V(wx + q)
  \E^{-x^2/2}\,\mathrm{d}x \approx \sqrt{2}\sum_{i=1}^n
  g_i V(\sqrt{2}wx_i + q),
\end{equation}
where $\{x_i\}_{i=1}^n$ and $\{g_i\}_{i=1}^n$ are the corresponding
nodes and weights. The second scheme is direct summation of a power
series, Eq.~\eqref{eq:Vint_exp}. It is straightforward when analytical
derivatives of $V(x)$ are known, but computationally less convenient
because high-order derivatives are rarely available \latin{ab initio}.
However, this scheme allowed us to cross-check all the formulas and
also implement evaluation of $U^1(q,w)$, Eq.~\eqref{eq:U_1}.

\subsection{\Acrlongpl{EOM}}
\label{sec:compl-model-spec}

The \gls{TDVP}-based dynamical scheme follows Eqs.~\eqref{eq:pq_dot}
and \eqref{eq:uw_dot} with the full $U(q, w)$ calculated by
Eqs.~\eqref{eq:Umorse_an} and \eqref{eq:Uammonia_an} for Morse and
ammonia double-well potentials, respectively. Phase evolution is
determined by Eqs.~\eqref{eq:lambda_dot} and \eqref{eq:Edef}.

In Heller's scheme \glspl{EOM} for the $(p, q)$ pair are purely
classical
\begin{subequations}
  \begin{align}
    \label{eq:heller_q_dot}
    {\dot q} = & \frac{p}{m}, \\
    \label{eq:heller_p_dot}
    {\dot p} = & -V'(q),
  \end{align}
\end{subequations}
while the \glspl{EOM} for the $(u, w)$ pair are Eqs.~\eqref{eq:w_dot}
and \eqref{eq:u_dot} with $U \equiv U^1(q,w)$ given by
Eq.~\eqref{eq:U_1}.

The extended semiclassical scheme uses Eqs.~\eqref{eq:pq_dot} and
\eqref{eq:uw_dot} with $U = U^1(q,w)$. In both approximate schemes the
phase is propagated by Eq.~\eqref{eq:lambda_dot} with $E$ evaluated by
Eq.~\eqref{eq:Edef_heller+ext_semicl}.

\subsection{Software}
\label{sec:software}

All the potentials and dynamical models as well as auxiliary routines
are implemented in \texttt{Julia} language~\cite{Julia:2017} and
available on GitHub~\cite{Ryabinkin:2024/github/tgwp}. The
\acrlongpl{EOM} are integrated using the explicit Runge--Kutta 5(4)
method with the updated tableau of
coefficients~\cite{Tsitouras:2011/camwa/770} available as
\texttt{Tsit5} algorithm in the
\texttt{DifferentialEquations.jl}~\cite{Rackauckas:2017/differentialequations}
Julia package.

\section{Results and discussion}
\label{sec:results-discussion}

\subsection{Convergence issues in approximations to the extended
  potential}
\label{sec:numer-appr-extend}

In general, numerical evaluation of the extended potential either by
integration or through summation of the Taylor series (see
Sec.~\ref{sec:eval-extend-potent}) becomes more challenging for larger
$w$. Additionally, Taylor series summation poses a problem of
convergence. Consider as an example the ammonia double-well model. The
width variable $w$ enters the exact expression for the extended
potential $U_\text{A}(q, w)$, Eq.~\eqref{eq:Uammonia_an}, as functions
$z^{-\frac{1}{2}}$ and $z^{-1}$ where $z = 1+2cw^2$. Taylor series of
both functions converge only for $|2cw^2| < 1$ (equivalently,
$|w| < \sqrt{\frac{1}{2c}}$), so is the Taylor
expansion~\eqref{eq:Vint_exp}.

We report convergence of numerical approximations (denoted as
$\widetilde U_\text{A}$) to $U_\text{A}(q, w)$, which employed the
10-point Gauss-Hermite quadrature and the Taylor series with 10 terms
in Table~\ref{tab:U_numerical_approx}. We show the mean and the
maximum of an absolute deviation
$|U_\text{A}(q,w) - \widetilde U_\text{A}(q,w)|$ for the range of $q$
from \SIrange[range-phrase={ to },range-units =
single]{-2.0}{2.0}{\bohr} (discretized on an equidistant grid with 101
points) and a few selected values of $w$: \SIlist[list-units =
single]{0.1;0.6;1.0}{\bohr}. For the smallest value of
$w = \SI{0.1}{\bohr}$ both numerical schemes converge to high
accuracy. For $w = \SI{0.6}{\bohr}$, which is close to the convergence
radius \SI{0.604}{\bohr}, the Taylor series demonstrates noticeable
deviations from reference values; it is also converges very slowly
upon adding more terms. Finally, for $w = \SI{1.0}{\bohr}$ the Taylor
series clearly diverge, while the Gauss-Hermite scheme is still
reasonably accurate. Gaussian integration is applicable for larger
values of $w$, but more points are needed for the same accuracy.
Larger integration grids, however, makes multi-dimensional
generalizations challenging, which may prompt researchers to consider
Monte Carlo integration instead.

It must be emphasized that convergence issues are independent of
representation: while they were demonstrated in
parameterization~\eqref{eq:1dgwp_uw_unnormalized}, they impact
\emph{any} approximate schemes that exploit the series expansion in a
variable related to the width.
\begin{table}
  \centering
  \caption{Convergence of numerical approximations
    $\widetilde U_\text{A}$ to the exact ammonia extended potential
    $U_\text{A}$.}
  \begin{minipage}{1.0\linewidth}
    \begin{tabularx}{1.0\linewidth}{@{}Xcc@{}} 
      \toprule %
      Approximation   & \multicolumn{2}{X}{\centering $|U_\text{A}(q, w) - \widetilde U_\text{A}(q, w)|$, \si{\hartree}} \\
      \cmidrule{2-3}
                      & mean  & max \\
      \midrule
      \multicolumn{3}{c}{$w=\SI{0.1}{\bohr}$} \\
      Gauss--Hermite & \num{2.9e-17} & \num{8.3e-17} \\
      Taylor & \num{1.8e-12} & \num{5.3e-12} \\
      \multicolumn{3}{c}{$w=\SI{0.6}{\bohr}$} \\
      Gauss--Hermite & \num{3.2e-7} & \num{8.2e-7} \\
      Taylor & \num{0.0021} & \num{0.0062} \\
      \multicolumn{3}{c}{$w=\SI{1.0}{\bohr}$} \\
      Gauss--Hermite & \num{8.8e-5} & \num{0.00019} \\
      Taylor & \num{0.53} & \num{1.5} \\
      \bottomrule
    \end{tabularx}
  \end{minipage}
  \label{tab:U_numerical_approx}
\end{table}

\subsection{Static limit}
\label{sec:static-limit}

\begin{figure}
  \centering \includegraphics[width=1.0\linewidth]{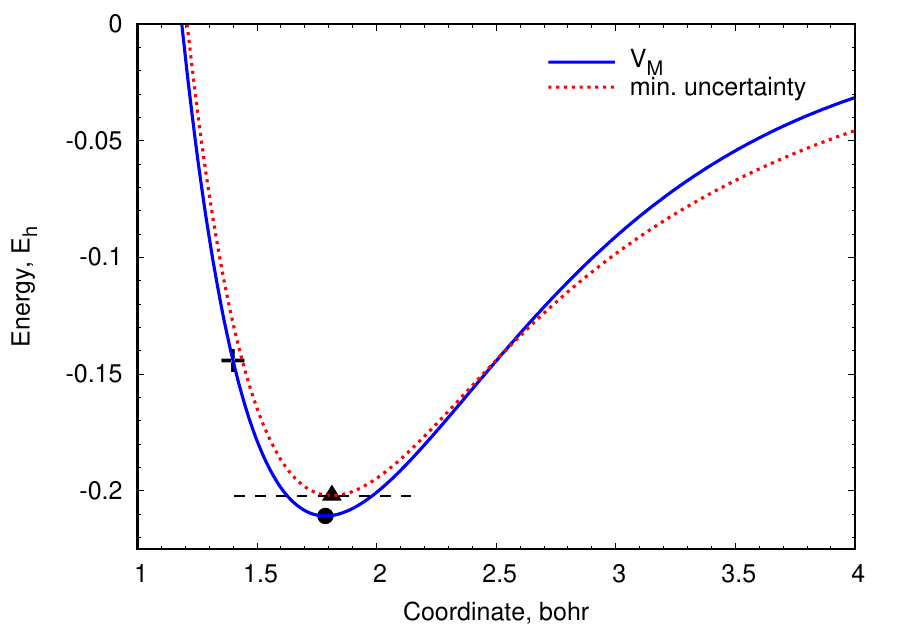}
  \caption{Quantum Morse potential $V_\text{M}$ (solid line) and the
    minimum uncertainty potential $\min_w U_\text{M}(q, w)$ (dotted
    line). A dashed horizontal line is a position of the ground
    vibrational level of $V_\text{M}$. A dot labels the (global)
    minimum of $V_\text{M}$ whereas a triangle is a coordinate of the
    global minimum of $U_\text{M}(q, w)$. A black cross marks the
    initial position of a wave packet for the setup discussed in
    Sec.~\ref{sec:extend-semicl-model}.}
  \label{fig:morse_pots}
\end{figure}

Consider for simplicity the Morse potential, which has only one
(global) minimum. When the amplitude of oscillations
vanishes~\footnote{Vanishing amplitudes may be a result of statistical
  averaging of multiple trajectories with random initial conditions,
  so the static limit is important when dynamical simulations are used
  to compute thermodynamic properties.}, one obtains a static limit,
in which all conjugate momenta are zero:
\begin{align}
  \label{eq:static_limit_p}
  \dot p & = 0, \\
  \label{eq:static_limit_u}
  \dot u & = 0.
\end{align}
For the Heller, the extended semiclassical, and the \gls{TDVP}-based
schemes the condition~\eqref{eq:static_limit_p} translates into
\begin{align}
  \label{eq:heller_static}
  \frac{\mathrm{d}V_\text{M}(x)}{\mathrm{d}x} & = 0, \\
  \label{eq:ext_semicl_static}
  \frac{\partial U_\text{M}^1}{\partial q} & = 0, \\
  \label{eq:tdvp_static}
  \frac{\partial U_\text{M}}{\partial q} & = 0,
\end{align}
respectively, where $U_\text{M}^1(q, w)$ is given by
Eq.~\eqref{eq:U_1} with $V \equiv V_\text{M}$. Thus, in Heller's
scheme a wave packet is stationary at the minimum (usually referred to
as $r_e$) of the quantum potential $V_\text{M}$, while for the other
models stationary points are different. Analytical consideration is
difficult for the full $U_\text{M}$, but is possible for
$U_\text{M}^1$. Solving Eq.~\eqref{eq:ext_semicl_static} one obtains
\begin{equation}
  \label{eq:U1_minimum_Morse}
  \E^{-b(q_\text{min}-a)} = 1 - \frac{3w^2b^2}{2+4w^2b^2} \equiv
  \alpha < 1\ \forall w>0,
\end{equation}
which means $(q_\text{min} - a) = -\ln{(\alpha)}/b > 0$, or
$q_\text{min} > a \equiv r_e$. We expect the same inequality holds for
the full $U_\text{M}(q,w)$, at least if anharmonicity is not too
strong. Turning to the numerical methods we located global minima of
$U_\text{M}$ and $U^1_\text{M}$ and found $q_\text{min}$ to be
\SI{1.8130}{\bohr} and \SI{1.8133}{\bohr}, respectively,---larger than
that for the Morse potential, whose
$q_\text{min} = r_e \equiv a = \SI{1.7858}{\bohr}$ (see
Table~\ref{tab:quantum_potentials_params} and
Fig.~\ref{fig:morse_pots}). Thus, Heller's scheme predicts noticeable
shorter equilibrium \ce{O-H} distance than the \gls{TDVP}-based or the
extended semiclassical scheme. This result is in agreement with
observations made in Ref.~\citenum{HyeonDeuk:2009/jcp/064501}, where
similar elongation was reported in simulations of liquid water.

Ranking of schemes changes if we consider how well the ground
vibrational energy is approximated. The full \gls{TDVP}-based scheme
conforms with the (stationary) variational principle approaching the
true energy from above. In the static limit, when
Eqs.~\eqref{eq:static_limit_p} and \eqref{eq:static_limit_u} hold,
finding the minimum of $U(q, w)$ is equivalent to minimization of the
expectation value of the quantum Hamiltonian [see
Eq.~\eqref{eq:H_func}] using a Gaussian as a wave-function
\textit{Ansatz}. The minimum of $U_\text{M}$ lies
\SI{1909.4}{\per\centi\meter} above the bottom of the Morse potential
well, which is within \SI{10}{\per\centi\meter} from the true
ground-state energy, \SI{1899.45}{\per\centi\meter}; see
Fig.~\ref{fig:morse_pots}. For the extended semiclassical scheme one
should seek for a minimum of $U^1_\text{M}$, which is located
\SI{1874.3}{\per\centi\meter} above the bottom of the potential well.
This scheme, therefore, underestimates the ground-state energy by
approximately \SI{25}{\per\centi\meter}. Finally, for Heller's scheme,
while the \emph{location} of the static wave packet can be found---in
$q$ variable it coincides with the minimum of $V_\text{M}$ whereas for
$w$ it is given by the condition
$\partial U_\text{M}^1/\partial w = 0$---its energy cannot be
unambiguously assigned. Measuring energy through $U_\text{M}$, or,
equivalently as an expectation value of the quantum Hamiltonian on the
static wave packet, one finds it equals to
\SI{1954.7}{\per\centi\meter}. Measured as a value of $U^1_\text{M}$
at the location of the static wave packet, the energy equals to
\SI{1919.4}{\per\centi\meter}, being approximately
\SI{20}{\per\centi\meter} above the exact. This value ranks Heller's
scheme second in accuracy. This situation can be explained by partial
error cancellation between approximations to the extended potential
and the variational principle for Heller's scheme. It must be admitted
that discrepancies between schemes are small compared to the absolute
energy of the ground state but they clearly demonstrate yet another
ambiguity existing in Heller's scheme.

For quantum potentials $V$ with a unique minimum (\latin{e.g.}
$V_\text{M}$), the difference between it and that of $U_\text{M}$ can
be interpret as \gls{ZPE}~\cite{Alimi:1992/jcp/2034}, see
Fig.~\ref{fig:morse_pots}. Even more, one can give a local,
position-dependent definition of \gls{ZPE} as follows. If we minimize
$U_\text{M}(q,w)$ with respect to $w$ at every $q$, then this quantity
is larger than $V_\text{M}(q)$ near the minimum of $V_\text{M}$ (the
difference is positive), but \emph{smaller} than $V_\text{M}(q)$ near
dissociation (the difference is negative); see the dotted curve in
Fig.~\ref{fig:morse_pots}. Therefore, the point-wise positivity of
thus-defined \gls{ZPE} is not guaranteed. As follows from
Eq.~\eqref{eq:U_1} this takes place approximately when $V''(x) < 0$.
However, this definition provides a simple explanation of tunneling:
near the top of a barrier where $V''(x) < 0$, \gls{ZPE} may act as a
source of energy driving barrier penetration. We elaborate on that in
Sec.~\ref{sec:tunneling-dynamics}.

The dotted line in Fig.~\ref{fig:morse_pots} depicts the minimal
possible energy of a wave packet, which ascents from the minimum
towards the dissociation infinitely slowly (adiabatically). One can
call such a packet the \emph{minimum uncertaity} wave packet, thus
explaining the naming of the corresponding potential curve.

\subsection{Dynamics in a small-amplitude regime}
\label{sec:small-ampl-regime}

Let us again concentrate on the Morse potential. We simulate dynamics
with the following initial conditions: the initial position
$q_0 = \SI{1.6}{\bohr}$, the initial width $w_0 = \SI{0.125}{\bohr}$,
all conjugated momenta $p_0$ and $u_0$, as well as the initial value
of the phase $\lambda_0$, are zero. Thus, the wave packet is initially
placed on a repulsive branch of the potential close to the minimum
(see Fig.~\ref{fig:morse_pots}) and its starting width is close yet
not exactly matches the coherent state width,
Eq.~\eqref{eq:coherent_state_width}, to induce small oscillations in
both $q$ and $w$ directions.

Since the adaptive algorithm to integrate \glspl{EOM} was used, the
integration time step was variable, but the solution was interpolated
on a regular time grid with a step size $dt = \SI{4.03}{\au}$ (atomic
units of time) using the default implementation available in
\texttt{DifferentialEquations.jl}. The total propagation time
corresponded to $n_t = 2^{14} = 16384$ time steps to provide
reasonable spectral resolution. Absolute and relative accuracy
requirements for integration were: \texttt{abstol} = \texttt{reltol} =
\num{1.0e-8}.

\subsubsection{Mechanical motion}
\label{sec:mechanical-motion}

\begin{figure}
  \centering \includegraphics[width=1.0\linewidth]{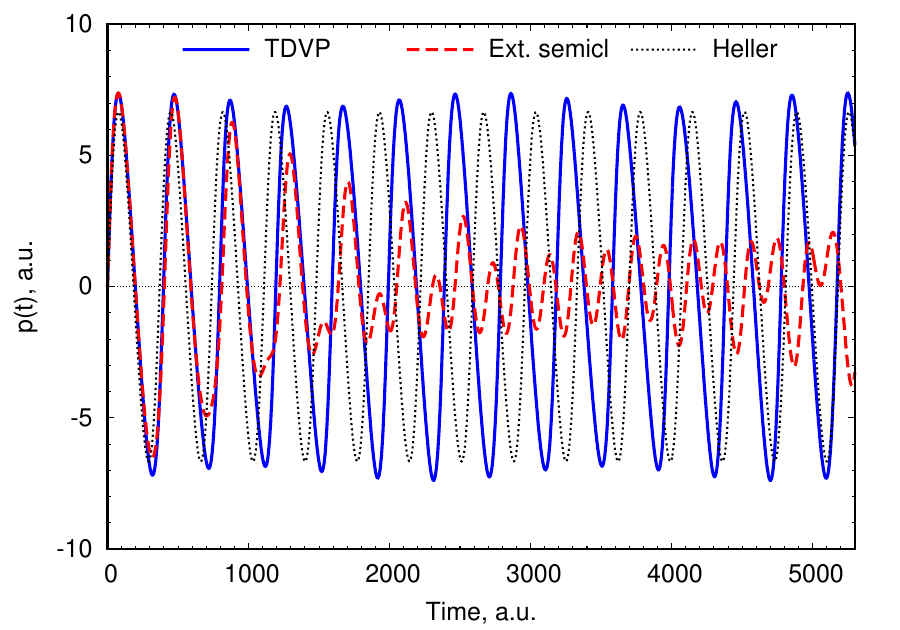}
  \caption{Linear momentum dynamics for the small-amplitude motion in
    the Morse potential.}
  \label{fig:momentum_dyn}
\end{figure}

Figure~\ref{fig:momentum_dyn} shows oscillations of a linear momentum
$p(t)$ of a wave packet. The most notable feature in
Fig.~\ref{fig:momentum_dyn} is a different amplitude of oscillations
in the extended semiclassical scheme as compared to both the
\gls{TDVP}-based and Heller's ones. We attribute these differences to
exaggerated coupling between position and width in the extended
semiclassical scheme that allows for faster energy exchange between
these \glspl{DOF}. However, a more important characteristic of
dynamics is an oscillation period. Simple counting reveals that in the
time span shown in Fig.~\ref{fig:momentum_dyn} there are 13
oscillation periods for the \gls{TDVP}-based scheme but 14 periods for
Heller's one. Therefore, the characteristic vibrational frequency in
the latter is 1/13 times higher than in former, which is substantial.
It is known that the red-shift of the \ce{OH} fundamental frequency is
an important quantum effect in liquid
water~\cite{HyeonDeuk:2009/jcp/064501}. Hence, Heller's scheme fails
to properly account for this quantum anharmonic effect.

\begin{figure}
  \centering \includegraphics[width=1.0\linewidth]{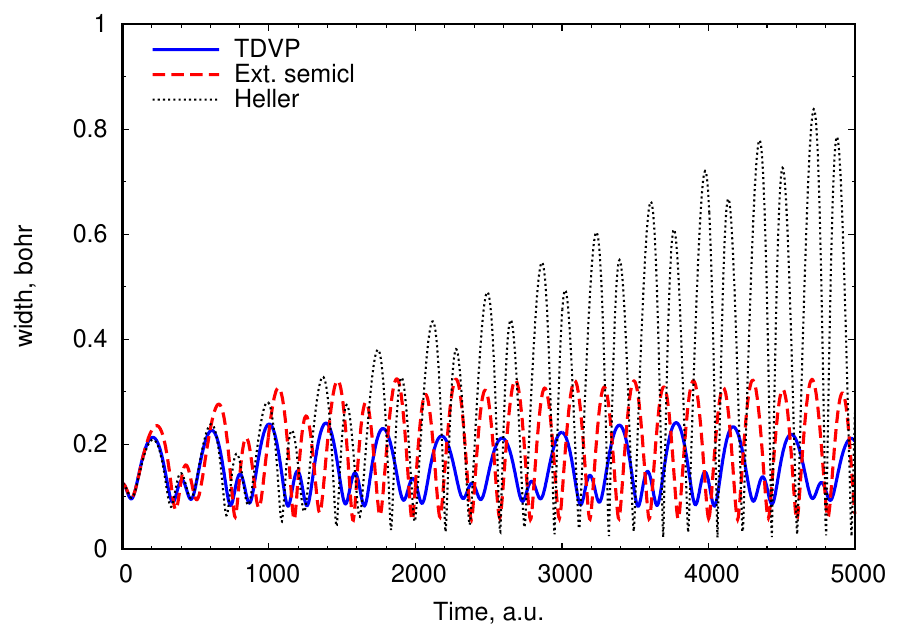}
  \caption{Width dynamics for the small-amplitude motion in the Morse
    potential.}
  \label{fig:width_dynamics}
\end{figure}

To our great surprise, we discovered a much more fundamental flaw of
Heller's scheme in describing the mechanical motion in this system,
which is clearly visible in Fig.~\ref{fig:width_dynamics}. Width
oscillations remain finite in the \gls{TDVP}-based and extended
semiclassical schemes, but in Heller's scheme grow linearly in time
without bounds. As follows from Eqs.~\eqref{eq:heller_q_dot} and
\eqref{eq:heller_p_dot} the width dynamics in Heller's scheme does not
affect linear motion of a wave packet. On the contrary, the
\glspl{EOM} for the $(u,w)$ pair, Eqs.~\eqref{eq:w_dot} and
\eqref{eq:u_dot} carry parametric dependence on the position variable
$q$. Thus, the observed phenomenon can be considered as a form of a
\emph{parametric resonance}.

A natural question is whether this phenomenon is universal or specific
to the Morse potential.
We ran a small-amplitude dynamics for the ammonia potential; results
are shown in Appendix~\ref{sec:width-dynam-ammon}. We observe very
similar behaviour of the width for Heller's dynamics: an unbound
oscillatory growth of the amplitude albeit on a different time scale.
Thus, the unbound growth of the width amplitude has nothing to do with
natural spreading of a wave packet in regions where a potential has
vanishing second- and higher-order derivatives (``flat
regions'')---the ammonia potential is nowhere flat. The unbound growth
does not exist for strictly quadratic potentials, but even minuscule
anharmonicity causes it to manifest. Fundamentally, this phenomenon is
due to the lack of back-reaction of the width dynamics onto position
of a Gaussian wave packet. Re-establishing of this coupling, as was
done in the extended semiclassical scheme is therefore crucial for
curing this deficiency; otherwise, an evolving Gaussian wave packet
must not be used to compute quantum-mechanical averages of any
property. In light of this prediction, it is interesting to analyse
(power) spectra predicted by the schemes.

\subsubsection{Spectra}
\label{sec:spectra}

We computed spectra via the Fourier transform of the autocorrelation
function, Eq.~\eqref{eq:autocorrelation_function}, as
\begin{equation}
  \label{eq:spectrum}
  I(\omega) = \lim_{T\to\infty}{\frac{1}{2\pi}\int\limits_{-T}^T C(t) \E^{\I(\omega -
      E_0/\hbar)t}\mathrm{d}t},
\end{equation}
where $E_0$ is the origin of the energy scale, which is chosen to be
the energy of the bottom of the Morse potential well in order to
enable comparison of excitation energies with eigenenergies.

\begin{figure}
  \centering \includegraphics[width=1.0\linewidth]{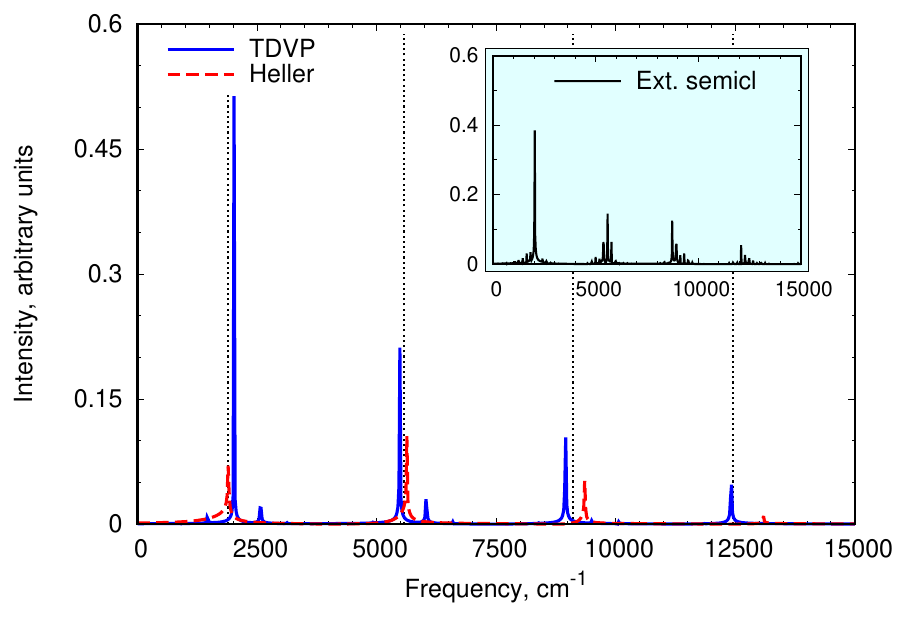}
  \caption{Absorption spectra $|I(\omega)|$ computed by
    Eq.~\eqref{eq:spectrum} (arbitrary units). Vertical black dotted
    lines are vibrational eigenlevels of the Morse potential.}
  \label{fig:spectra}
\end{figure}

Numerically, the integral in Eq.~\eqref{eq:spectrum} is evaluated from
$0$ to finite $T$ using the fact that $C(-t) = C^{*}(t)$ (time
reversibility) via the Fast Fourier transform and taking the real part
of the result. Truncation at finite $T$, however, may introduce a jump
discontinuity, unless the interval $[0; T)$ contains an integer
multiple of an oscillation period. For such periodic signals the
imaginary part of $I(\omega)$ vanishes and one could use either real
or absolute parts. We carefully chosen the upper limit
$T = t_0 + (n_t-1)dt = \SI{66023.49}{\au}$ (see
Sec.~\ref{sec:small-ampl-regime}) to minimize the jump
discontinuity~\footnote{Another popular approach to minimize
  truncation effects is damping of the autocorrelation function by
  multiplying it by $\exp{(-\kappa t)}$ with $\kappa > 0$ to ensure
  $C(T) \to 0$ along with all its derivatives. This corresponds to a
  convolution of the spectrum with a Lorentzian line shape function
  with the width $\sim 1/\kappa$. We do not apply this technique
  here.}. The absolute value $|I(\omega)|$ (in arbitrary units) along
with the vibrational eigenlevels of the Morse potential (vertical
black lines) are plotted in Fig.~\ref{fig:spectra}.

As evident from Fig.~\ref{fig:spectra}, the \gls{TDVP}-based scheme
shows peaks at energies that are very close to vibrational
eigenenergies. Moreover, relative peak intensities are reproduced
fairly well: expansion of the initial Gaussian over a set of
bound-state Morse eigenfunctions gives relative weights
1:0.59:0.26:0.11, which are visually close to majors peaks heights in
Fig.~\ref{fig:spectra}. On the other hand, the \gls{TDVP}-based scheme
also demonstrates (spurious) weak satellite peaks (\latin{e.g.} one
near \SI{6000}{\per\centi\meter}), which are probably caused by a
resonance between position and width oscillations. Satellite peaks are
more pronounced in the extended semiclassical scheme (see the inset in
Fig.~\ref{fig:spectra}), but the positions and intensities of the
major peaks are reasonably close to those given by the
\gls{TDVP}-based counterpart.

Despite the pathological \textit{mechanical} behaviour discussed in
Sec.~\ref{sec:mechanical-motion}, the spectrum given by Heller's
scheme is surprisingly accurate. The ground- and the first
excited-state peak positions are very close to the exact energies,
only the second and higher peaks start to deviate. It appears,
therefore, that the spurious large-amplitude width oscillations have
little impact on the computed autocorrelation function and hence, the
spectrum. We explain this by the observation that spectra are mainly
determined by a short-time behaviour of the autocorrelation function
$C(t)$, Eq.~\eqref{eq:autocorrelation_function}. In practice, it is
usually damped to zero to account for the environmental effects, and
the unbound growth of the width provides a somewhat strange mechanism
to drive $C(t)$ to zero. On the other hand, the \emph{width} of the
spectral lines, which is inversely proportional to a natural lifetime
of a wave packet, in this case has nothing to do with any
environmental effects, and in fact is determined by the strength of
anharmonicity of the parent quantum potential.

Spectral intensities are less satisfactory reproduced by Heller's
scheme. For example, the most intensive (highest) peak is assigned to
the first excited state, not the ground one. Nevertheless, in typical
spectroscopic applications intensities are usually considered to be of
lesser importance than peaks positions.

Overall, our findings partly explain why the flaws of Heller's scheme
went unnoticed in the majority of previous studies, which were mainly
focused on spectroscopic applications. In one
work~\cite{Wehrle:2015/jpca/5685}, however, rising amplitude of the
semiclassical energy~\eqref{eq:Edef_heller+ext_semicl} in a course of
dynamics was reported---see Fig.~3 there. Ultimately, this phenomenon
is related to the pathological width dynamics.

\subsection{Tunneling dynamics}
\label{sec:tunneling-dynamics}

\begin{figure}
  \centering %
  \includegraphics[width=1.0\linewidth]{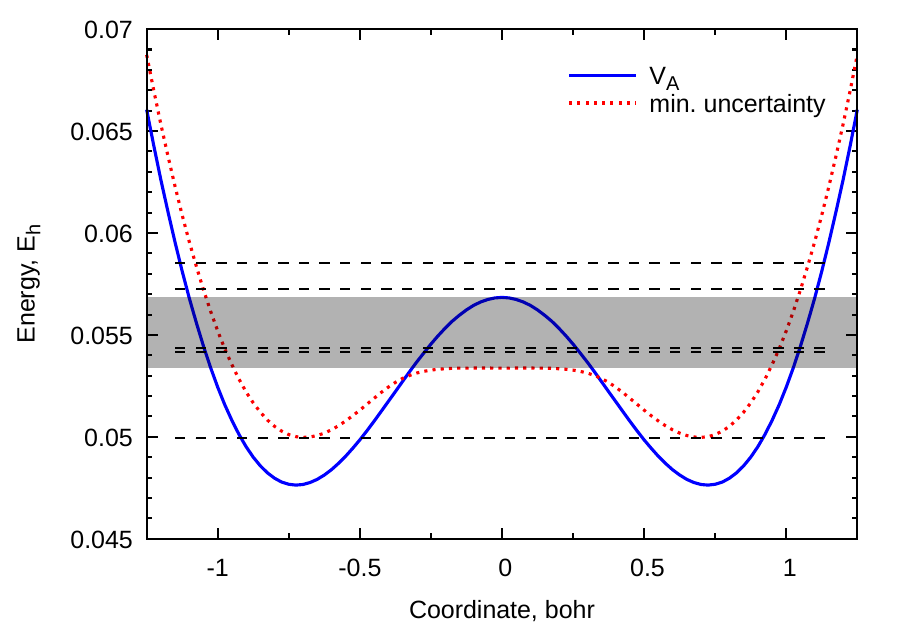}
  \caption{Ammonia double-well potential $V_\text{A}$ (solid line) and
    $\min_w U_\text{A}(q, w)$ (dotted line). Dashed horizontal lines
    denote vibrational levels of $V_\text{A}$. Note that the ground
    and the first excited state are almost degenerate and visually
    indiscernible. A shaded area is a region of energies in which
    tunneling of a wave packet is possible in the \gls{TDVP}-based
    scheme.}
  \label{fig:ammonia_pots}
\end{figure}

Tunneling dynamics is modelled for the ammonia double-well potential
shown in Fig.~\ref{fig:ammonia_pots}.
As it was done for the Morse potential, we numerically computed the
minimal uncertainty potential $\min_w U_\text{A}(q,w)$ shown as a red
dotted line in Fig.~\ref{fig:ammonia_pots}. In a range of energies
between its local maxima and the maximum of $V_\text{A}$ near $q=0$
(shaded area in Fig.~\ref{fig:ammonia_pots}) a wave packet can tunnel
through a barrier in the \gls{TDVP}-based scheme. We simulated the
dynamics in the tunneling regime using the following initial
conditions: initial position and width are \SI{-1.0}{\bohr} (the left
well) and \SI{0.15}{\bohr}, respectively, all the conjugate momenta
and the initial phase are zero. The accuracy parameters for the time
integration, \texttt{abstol} and \texttt{reltol}, were tightened to
\num{1.0e-12}. Trajectory data were interpolated on a regular time
grid with a time step $dt = \SI{4.03}{\au}$.

\begin{figure}
  \centering \includegraphics[width=1.0\linewidth]{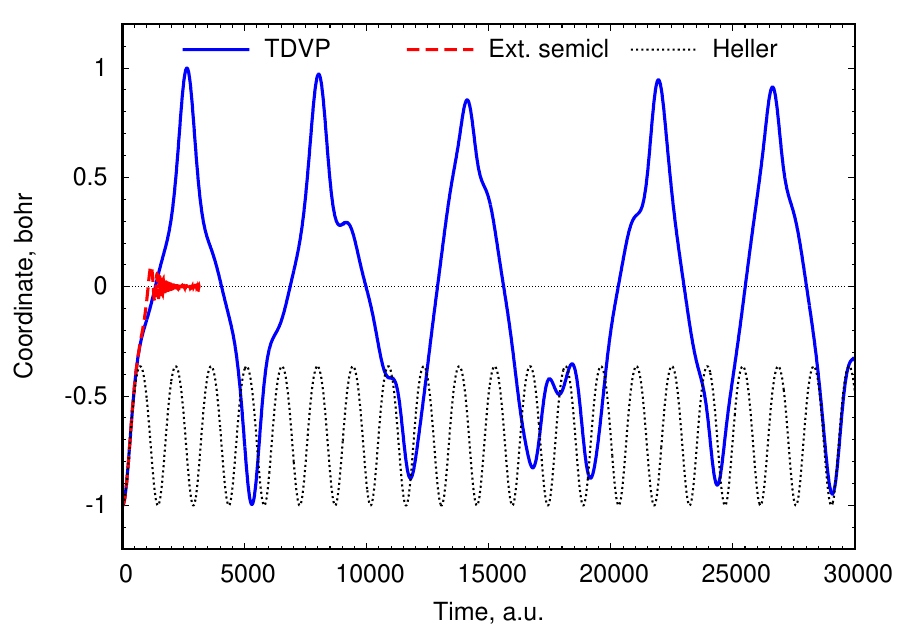}
  \caption{Time evolution of the coordinate of a center of Gaussian
    wave packet $q(t)$ in the double-well ammonia potential; see
    Fig.~\ref{fig:ammonia_pots}.}
  \label{fig:ammonia_tunnel_traj}
\end{figure}

Time evolution of the coordinate of a center of a wave packet $q(t)$
is shown in Fig.~\ref{fig:ammonia_tunnel_traj}. As anticipated, the
wave packet tunnels through a barrier in the \gls{TDVP}-based scheme
while in Heller's scheme it stays inside the left well. However, the
most notable feature in Fig.~\ref{fig:ammonia_tunnel_traj} is the
trapping of the semiclassical trajectory at the top of the barrier,
$q=0$. This unexpected result can be explained as follows. The
effective potential in the extended semiclassical scheme
$U^1_\text{A}(q,w)$, Eq.~\eqref{eq:U_1}, contains a contribution that
depends on the curvature of the potential $V_\text{A}''(q)$ at a
position of a wave packet. Near $q=0$ the curvature is
\emph{negative}, so both $V(q)$ and $V''(q) w^2/2$ terms have a shape
of a curved-down parabola. If a wave packet passes this region slowly
enough, these contributions drive the width of the packet to infinity
and essentially remove the maximum at $q=0$ turning it into a minimum.
It must be noted that a very shallow parabolic well centered at $q=0$
with the depth $\sim \SI{2e-6}{\hartree}$ exists for the minimum
uncertainty potential $\min_w U_\text{A}(q,w)$ even for the exact
$U_\text{A}$ though this region appears essentially flat in
Fig.~\ref{fig:ammonia_pots}. Thus, the wave packet may be trapped in
the \gls{TDVP}-based scheme too, but this requires fine tuning of the
initial parameters: the packet must be placed on the top of the
barrier from the beginning, and its width must be carefully optimized.
No divergence of the width to infinity is possible, so one may
speculate that the \gls{TDVP}-based scheme predicts some sort of
quantum \emph{resonance} in this case.

We validated our explanation by performing two tests. First, we
verified that a sufficiently fast wave packet could avoid trapping. If
the initial position is set to $q_0 = \SI{-1.6}{\bohr}$, which
corresponds to the initial energy well above the top of the barrier
(see Fig.~\ref{fig:ammonia_pots}), then the center of a wave packet
exhibits similar dynamics in all three schemes for at least
\SI{1.3e5}{\au}. Second, a slow wave packet in the region of negative
curvature can be realized near a classical turning point of the
\emph{Morse} potential in a region where $V_\text{M}''(q) < 0$. We
consider this scenario in the next section.

\subsection{Trapping of the extended semiclassical trajectory in the
  large-amplitude motion in the Morse potential well}
\label{sec:extend-semicl-model}

We simulated large-amplitude dynamics for the Morse potential using
the following initial parameters: the initial position and width are
\SI{1.4}{\bohr} and \SI{0.125}{\bohr}, respectively, all the
conjugated momenta and the initial phase are zero. These initial
conditions are labelled by a black cross in Fig.~\ref{fig:morse_pots}.
The accuracy parameters for the time integration, \texttt{abstol} and
\texttt{reltol}, were set to \num{1.0e-12}, the time step $dt$ for
interpolation was, as before, \SI{4.03}{\au}.
\begin{figure}
  \centering \includegraphics[width=1.0\linewidth]{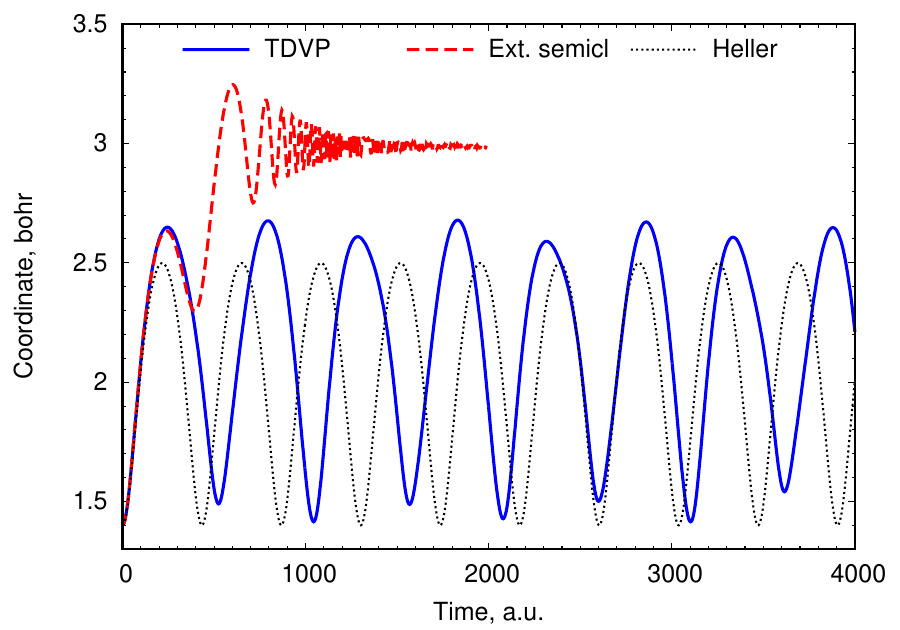}
  \caption{Same as in Fig.~\ref{fig:ammonia_tunnel_traj} but for the
    Morse potential. The initial wave packet position is marked with a
    black cross in Fig.~\ref{fig:morse_pots}.}
  \label{fig:morse_lam}
\end{figure}

As anticipated, the trajectory given by the extended semiclassical
scheme, gets trapped after initially entering into the negative
curvature region of $V_\text{M}$. Only when the initial energy is high
enough and above the dissociation limit, (\latin{e.g} for
$q_0 = \SI{1.1}{\bohr}$), the extended semiclassical dynamics avoids
trapping. That is, the extended semiclassical dynamics beaks down near
classical turning points in regions of negative curvature of the
quantum potential.

\section{Conclusions}
\label{sec:conclusions}

The main message of our work is that the \gls{TDVP}-based propagation
scheme is devoid of physical artefacts and only it should be used to
simulate short- and intermediate-time quantum dynamics with anharmonic
potentials. It is able to capture important quantum effects. It
properly accounts for \gls{ZPE} both in a static picture (see
Sec.~\ref{sec:static-limit}) by distinguishing the (ro-)vibrational
state-averaged equilibrium distance $r_0$ and $r_e$ and providing
$r_0 > r_e$~\cite{Hirano:2021/jms/1229}, and in dynamics by predicting
the red shift of a frequency of mechanical oscillations (see
Sec.~\ref{sec:mechanical-motion}). It provides accurate spectral lines
positions and intensities if the initial energy is not too high; see
Fig.~\ref{fig:spectra}. Finally, it describes quantum tunneling
\latin{via} the mechanism that can be dubbed as \gls{ZPE}-assisted: in
regions of negative potential curvature energy ``stored'' in the width
variable can be used by a wave packet to overcome a barrier; see
Sec.~\ref{sec:tunneling-dynamics}. The missing quantum effects are
related to inability of a single Gaussian to bifurcate and
(self)-interfere. Additionally, experimental initial conditions are
better represented by a thermal density matrix rather than a pure wave
function. However, a satisfactory density-based \gls{TDVP} formalism
was only recently developed~\cite{Picconi:2019/jcp/224106} on the
basis of the matrix Schr\"odinger equation for open
systems~\cite{JoubertDoriol:2014/jcp/234112}; to the best of our
knowledge it is not yet been applied to the thawed Gaussian dynamics.

The oldest and by far the most widely used Heller scheme gained its
popularity due to relative algorithmic simplicity and compatibility
with on-the-fly calculations of \latin{ab initio} energies, gradients,
and Hessian (force constant) matrices---there is no need to fit a
\acrlong{PES} beforehand. Additionally, the scheme provides very easy
access to vibronic spectra via the directly simulated autocorrelation
function $C(t)$, Eq.~\eqref{eq:autocorrelation_function}. We found,
however, that it failed to provide a better physical picture than that
given by classical mechanics. Because the \acrlongpl{EOM} for this
scheme, Eqs.~\eqref{eq:heller_q_dot} and \eqref{eq:heller_p_dot}, are
purely classical, a Gaussian wave packet visits exactly the same
regions of a $(p,q)$ phase space as a classical trajectory does.
Heller's scheme does not properly account for quantum anharmonic
effects. Namely, it does not distinguish between $r_0$ and $r_e$ in
the static limit and cannot predict a red shift of a frequency of
mechanical oscillations, see Secs.~\ref{sec:static-limit}
and~\ref{sec:mechanical-motion}. Moreover, due to the lack of coupling
between position and width, Heller's dynamics is unable to describe
tunneling \latin{via} the \gls{ZPE}-assisted mechanism. The most
serious drawback of Heller's scheme is unbound oscillatory growth of
the width in a process that resembles a parametric resonance as
demonstrated in Sec.~\ref{sec:mechanical-motion}. This phenomenon has
nothing to do with proper wave packet spreading on flat potentials. It
does not exist for strictly quadratic potentials and appears only for
anharmonic ones with non-vanishing higher derivatives. In view of all
these fundamental issues, it might be not clear though, why absorption
spectra predicted by Heller's scheme are often so surprisingly
accurate. We explain this as follows. In many cases spectra are
predominantly determined by a short-time behaviour of the
autocorrelation function $C(t)$. In practice, $C(t)$ is usually damped
to zero in account for environmental effects, and the unbound growth
of the width provides a somewhat strange mechanism to drive $C(t)$ to
zero. On the other hand, the width of spectral lines, which is
inversely proportional to a natural lifetime of a wave packet, in this
case has nothing to do with any environmental effects, and in fact is
determined by the strength of anharmonicity of the parent quantum
potential. Summarising, we strongly advise researchers against any use
of a Gaussian wave packet parametrized by results of Heller's dynamics
in computations of expectation values of position- and especially
width-dependent operators for times that are greater than a few width
oscillation periods---see, for example, Fig.~\ref{fig:width_dynamics}.

The extended semiclassical scheme was devised to improve upon the
Heller counterpart. While it indeed corrects the dynamics in the
static limit and for small-amplitude motions, it is susceptible to
completely unphysical trapping of a trajectory in regions of negative
potential curvature when a wave packet moves slowly. That means that
neither tunneling (see Sec.~\ref{sec:tunneling-dynamics}), nor
large-amplitude motion (see Sec.~\ref{sec:extend-semicl-model}) can be
described by this scheme without modifications. One needs to go to
higher-order terms in Eq.~\eqref{eq:Vint_exp} to reach a derivative
that is positive in the region where a wave packet evolves, but in
this case the problem of convergence of the Taylor series illustrated
in Sec.~\ref{sec:numer-appr-extend} may render this approach invalid,
not even mentioning the fact that higher than third derivatives are
virtually unavailable in electronic structure packages.

While the current manuscript was in preparation, we became aware of
recent works~\cite{Vanicek:2023/jcp/014114, Fereidani:2023/jcp/094114,
  Fereidani:2024/jcp/044113} that have a strong overlap with our work.
Our analysis is founded on the properties of the extended potential
$U$ and complements these works substantially identifying flaws that
have never been previously documented.

Reiterating, we strongly believe that the \gls{TDVP}-based scheme is
best suited for reliable simulations of approximate quantum dynamics.
The main challenge, however, is in combining it with \latin{ab initio}
potentials, preferably computed on the fly. As we shown in
Sec.~\ref{sec:numer-appr-extend}, despite its poor scaling, a
grid-based scheme may be exploited for low-dimensional problems. For
high-dimensional systems a prospective solution could be to fit
\latin{ab initio} potentials locally using non-parametric regression
techniques. The nuclear configurations used for the fitting could be
sampled by initially identifying the nuclear subspace most relevant
for dynamics. Indeed, as shown in our follow-up
paper~\cite{Gherib:2024/arxiv/2405.00193}, not all nuclear \glspl{DOF}
are always dynamically relevant, and many can be neglected in fitting.
Sampling along a small number of dynamically relevant \glspl{DOF}
might mitigate the curse of dimensionality and ultimately allow the
\gls{TDVP}-based scheme to be utilised on industrially relevant
molecules.

\begin{acknowledgements}
  I.G.R. is grateful to Profs. P.~Brumer and A.F.~Izmaylov from the
  University of Toronto for bringing attention to the thawed Gaussian
  dynamics back in 2017.
\end{acknowledgements}

\appendix*

\section{Width dynamics in the ammonia potential}
\label{sec:width-dynam-ammon}

\begin{figure}[!b]
  \centering
  \includegraphics[width=0.95\linewidth]{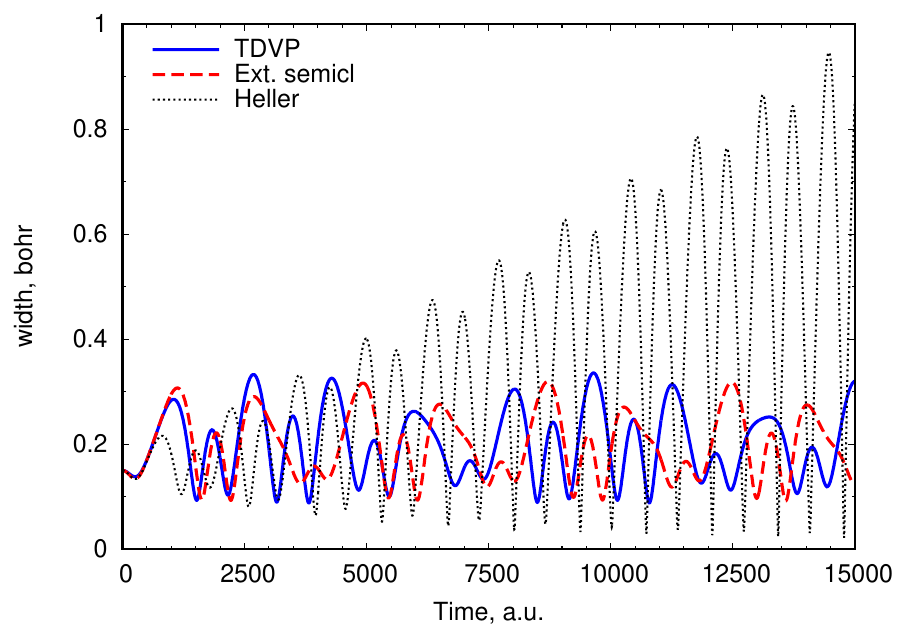}
  \caption{Width dynamics for the ammonia model potential.}
  \label{fig:ammonia_sam_dyn}
\end{figure}

We initiate small-amplitude dynamics near the left minimum of the
ammonia potential (see Fig.~\ref{fig:ammonia_pots}) using the
following initial conditions: the initial position
$q_0 = \SI{-0.9}{\bohr}$, the initial width $w_0 = \SI{0.15}{\bohr}$,
and $p_0 = u_0 = \lambda_0 = 0$. The width dynamics is shown in
Fig.~\ref{fig:ammonia_sam_dyn}. Contrary to the full \gls{TDVP} and
the extended semiclassical models which predict oscillations with
finite amplitudes, Heller's model exhibits linear growth of the
amplitude similar to that shown in Fig.~\ref{fig:width_dynamics}. This
clearly demonstrates that this phenomenon is not merely caused by the
existence of flat regions in the potential (as in the Morse
potential), but is a more general drawback of Heller's scheme.

\bibliography{paramres}

\end{document}

%% file: acronyms.tex
\newacronym[longplural={degrees of freedom}, firstplural={degrees of
  freedom (DOF)}, plural={DOF}]{DOF}{DOF}{degree of freedom} %
\newacronym[longplural={equations of motion}, firstplural={equations
  of motion (EOM)}, plural={EOM}]{EOM}{EOM}{equation of motion} %
\newacronym{TDSE}{TDSE}{time-dependent Schr\"odinger equation}
\newacronym{TIDSE}{TIDSE}{time-independent Schr\"odinger equation}
\newacronym{TDVP}{TDVP}{time-dependent variational principle} %
\newacronym{DFVP}{DFVP}{Dirac--Frenkel variational principle} %
\newacronym{MCTDH}{MCTDH}{multiconfguration time-dependent Hartree}
\newacronym{BCH}{BCH}{Baker--Campbell--Hausdorff}

\newacronym{vMCG}{vMCG}{variational multi-configurational Gaussian}
\newacronym{FMS}{FMS}{full multiple spawning} %
\newacronym{G-MCTDH}{G-MCTDH}{Gaussian-based multi-configuration time-dependent Hartree} %
\newacronym{MAE}{MAE}{mean absolute error} %
\newacronym{pMAER}{\%MAER}{percentage of mean absolute error reduction}

\newacronym{GWP}{GWP}{Gaussian wavepacket} %
\newacronym{TGWP}{TGWP}{thawed Gaussian wavepacket} %
\newacronym{GHA}{GHA}{global harmonic approximation} %
\newacronym{LHA}{LHA}{local harmonic approximation} %
\newacronym{ZPE}{ZPE}{zero-point energy} %
\newacronym{VM}{VM}{vibrational mode} %
\newacronym{CVM}{CVM}{curvilinear vibrational mode}

\newacronym{ML}{ML}{machine learning} %
\newacronym{ML-PES}{ML-PES}{machine-learned potential energy surface} %
\newacronym{KRR}{KRR}{kernel ridge regression} %
\newacronym{GPR}{GPR}{Gaussian process regression} %

\newacronym{OLED}{OLED}{organic light-emitting diode}
\newacronym{NISQ}{NISQ}{noisy intermediate-scale quantum}

\newacronym{JW}{JW}{Jordan--Wigner} %
\newacronym{BK}{BK}{Bravyi--Kitaev} %

\newacronym{QPE}{QPE}{quantum phase estimation} %
\newacronym{VQE}{VQE}{variational quantum eigensolver} %
\newacronym{QMF}{QMF}{qubit mean-field} %
\newacronym{QCC}{QCC}{qubit coupled cluster} %
\newacronym{iQCC}{iQCC}{iterative qubit coupled cluster} %
\newacronym{PQA}{PQA}{parametrized quantum annealing} %
\newacronym{DIS}{DIS}{direct interaction set} %
\newacronym[longplural={involutary linear combinations of
anti-commuting Paulis}, firstplural={involutory linear combinations of
anti-commuting Paulis (ILCAP)}, plural={ILCAP}]{ILCAP}{ILCAP}{involutory linear combination of anti-commuting Paulis} %

\newacronym{CAS}{CAS}{complete active space} %
\newacronym{PES}{PES}{potential energy surface} %
\newacronym{PEC}{PEC}{potential energy curve} %
\newacronym{AO}{AO}{atomic orbital} %
\newacronym{MO}{MO}{molecular orbital} %

\newacronym{CI}{CI}{configuration interaction} %
\newacronym{FCI}{FCI}{full configuration interaction} %
\newacronym{CASCI}{CASCI}{complete active space configuration interaction} %
\newacronym{MCSCF}{MCSCF}{multiconfigurational self-consistent
  field} %
\newacronym{CASSCF}{CASSCF}{complete active space self-consistent
  field} %
\newacronym{CC}{CC}{coupled cluster} %
\newacronym{UCC}{UCC}{unitary coupled cluster} %
\newacronym{UCCSD}{UCCSD}{unitary coupled cluster singles and
  doubles} %
\newacronym{CCSD}{CCSD}{coupled-cluster singles and doubles} %
\newacronym{CCSD-T}{CCSD(T)}{coupled-cluster singles and doubles and
  non-iterative triples} %
\newacronym{RHF}{RHF}{restricted Hartree--Fock} %
\newacronym{CIS}{CIS}{configuration-interaction singles} %
\newacronym{ROHF}{ROHF}{restricted open-shell Hartree--Fock} %
\newacronym{UHF}{UHF}{unrestricted Hartree--Fock} %
\newacronym{DMRG}{DMRG}{density-matrix renormalization group} %
\newacronym{DFT}{DFT}{density-functional theory} %
\newacronym{TDDFT}{TDDFT}{time-dependent density-functional theory} %
\newacronym{ENPT}{ENPT}{Epstein-Nesbet perturbation theory} %
\newacronym{MP}{MP}{M{\o}ller--Plesset perturbation theory} %
\newacronym{MP2}{MP2}{second-order M{\o}ller--Plesset perturbation theory} %
\newacronym{MRMP2}{MRMP2}{second-order multi-reference M{\o}ller--Plesset perturbation theory} %

\newacronym{SQP}{SQP}{sequential quadratic programming} %
\newacronym{MMA}{MMA}{method of moving asymptotes} %


%% file: operators.tex
\newcommand{\E}{\textrm{e}} %
\newcommand{\I}{\mathrm{i}\mkern1mu} %
\def\be{\begin{equation}} %
\def\ee{\end{equation}} %
\def\bea{\begin{eqnarray}} %
\def\eea{\end{eqnarray}} %


%% file: ms.bbl
\begin{thebibliography}{75}
\bibitem [{Notea()}]{Notea}%
  \BibitemOpen
  \bibinfo {note} {Vanishing amplitudes may be a result of statistical
  averaging of multiple trajectories with random initial conditions, so the
  static limit is important when dynamical simulations are used to compute
  thermodynamic properties.}\BibitemShut {Stop}%
\bibitem [{Noteb()}]{Noteb}%
  \BibitemOpen
  \bibinfo {note} {Another popular approach to minimize truncation effects is
  damping of the autocorrelation function by multiplying it by $\protect
  \qopname \relax o{exp}{(-\kappa t)}$ with $\kappa > 0$ to ensure $C(T) \to 0$
  along with all its derivatives. This corresponds to a convolution of the
  spectrum with a Lorentzian line shape function with the width $\sim 1/\kappa
  $. We do not apply this technique here.}\BibitemShut {Stop}%
\end{thebibliography}
